\documentclass[sigplan,10pt,nonacm,balance=false]{acmart}
\usepackage{amsmath,tabularx,array,graphicx}
\usepackage{placeins,stfloats}
\setcopyright{none}
\acmConference[]{}{}{}
\acmBooktitle{}
\acmDOI{}
\acmISBN{}
\hypersetup{hidelinks}
\graphicspath{{figures/}}
\newcommand{\code}[1]{\nolinkurl{#1}}
\newcommand{\system}{\mbox{CXLResume}}
\AtBeginDocument{\normalsize}

\title{CXLResume: Retained-State Checkpoint and Restore for In-Memory Services}
\author{Haonan Li}
\affiliation{\institution{}\city{}\country{}}
\email{honyli321@gmail.com}

\author{Zhaoxiang Huang}
\affiliation{\institution{}\city{}\country{}}
\email{zhaoxiang.huang@outlook.com}

\author{Zhihao Zhang}
\affiliation{\institution{}\city{}\country{}}
\email{zhihaoz6@foxmail.com}

\author{Yiming Zhang}
\authornote{Corresponding author.}
\affiliation{\institution{}\city{}\country{}}
\email{zhangyiming@cs.sjtu.edu.cn}

\renewcommand{\shortauthors}{Li et al.}
\date{}
\begin{document}
\begin{abstract}
Checkpointing and restoring an in-memory service typically require writing its memory pages to checkpoint images and refilling them at the destination. When application state already resides in shared memory that remains accessible after failover, copying those pages into and out of checkpoint images is unnecessary. We present \system, a prototype that retains the service's main heap objects in shared memory while checkpointing the remaining execution state and mapping metadata. Checkpointing preserves the shared heap at the same freeze point as the serialized process state. During recovery, \system{} remaps the same shared memory at its original virtual addresses before restoring execution, preserving direct pointers within and across shared and private memory. We evaluate \system{} with Redis and Memcached using QEMU CXL functional emulation. At 6 GiB, \system{} reduces combined checkpoint and restore time from 15.015 s to 0.242 s for Redis and from 19.770 s to 0.276 s for Memcached in the evaluated deployments. These results show that retaining shared application state removes large application heaps from the checkpoint data path and substantially reduces checkpoint and restore latency.
\end{abstract}
\maketitle
\section{Introduction}\label{sec:intro}
In-memory services must often move for maintenance, resource rebalancing, or host replacement while preserving their application state. Checkpoint/restore captures a running process and reconstructs it at a destination, but serializing and reloading a large heap can dominate service interruption. In our evaluation, checkpointing and restoring a Redis dataset with 6~GiB of values through local CRIU images takes 15.015~s. Shared checkpoint storage removes the need to transfer those images explicitly, yet application pages still pass through serialization and loading.

Shared CXL memory offers an opportunity to avoid this work: application objects can remain in place while execution moves. Memory-resident checkpoints, fabric-attached migration, and execution-environment reuse provide related mechanisms~\cite{vas,fam,trenv,cxlfork}. The problem we address is how to resume a service over its existing writable objects while preserving native references between shared objects and ordinary process memory.

We present \system{}, a prototype that places eligible allocations in CXL memory before migration and checkpoints the remaining execution state. Shared objects and captured process state correspond to the same freeze point. The destination reconstructs the original virtual-address mappings before resuming threads, and the source remains stopped during takeover. This arrangement preserves direct pointers without traversing the application's object model or deferring bulk heap retrieval until requests arrive.

We implement this design with shared allocation and a CRIU mapping plugin, and evaluate Redis and Memcached using DRAM-backed QEMU CXL emulation. Our comparison uses local checkpoint storage for CRIU and shared checkpoint storage for \system{}. At 6~GiB, \system{} completes Redis checkpoint and restore in 0.242~s. Continuous-request experiments connect this reduction to service resumption, while data-structure and trace-derived workloads examine continuation across different object layouts. The paper contributes the retained-object checkpoint design, its address-preserving implementation, and an evaluation of its effect on migration interruption.

\section{Background and motivation}\label{sec:background}
\subsection{Checkpointing and service interruption}
Process migration must preserve both the objects a service has accumulated and the execution that uses them. CRIU captures a process tree's memory layout, required page contents, thread contexts, and supported operating-system resources~\cite{criu}. The destination reconstructs these dependencies before the application continues. For an in-memory service, reconstructing a logical key/value dataset alone would leave stacks, allocator state, and live resource references outside that continuation path.

\begin{figure*}[tbp]
\centering\includegraphics[width=\linewidth]{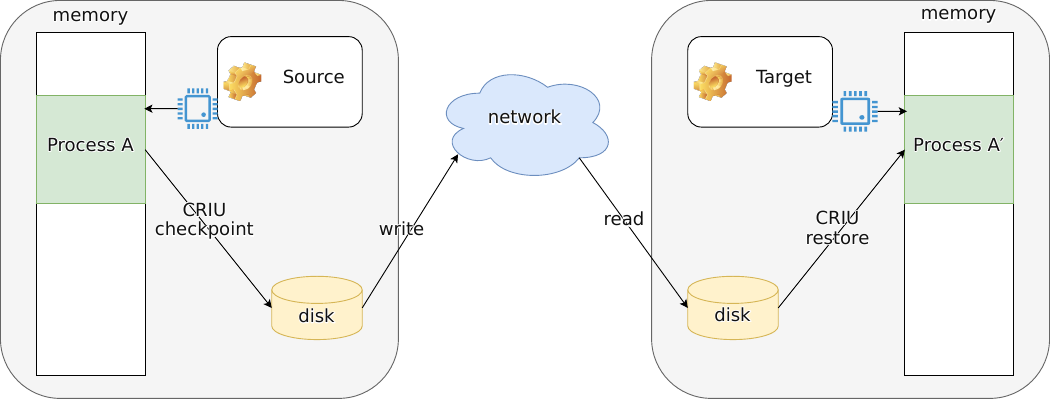}
\caption{Conventional CRIU checkpoint and restore. Application pages pass through checkpoint images before execution resumes. Our CRIU configuration writes local images and transfers them before restore.}
\label{fig:ordinary-path}
\Description{Source memory is serialized into checkpoint images and restored into a destination process.}
\end{figure*}

In a conventional checkpoint, required anonymous pages enter page images and are loaded into the destination address space during restore (Figure~\ref{fig:ordinary-path}). The storage location changes how these images become available. Local images must be transferred before the destination can use them. Shared checkpoint storage provides access to the same image files, but the source still serializes the required pages and the destination still loads them. Shared image access therefore leaves the application's bulk memory on the checkpoint path.

Pre-copy and post-copy change when memory transfer occurs relative to execution~\cite{clark,m3u}. Pre-copy transfers pages while the source runs and captures the remaining changes at handoff; post-copy allows execution to resume while some memory is still being fetched. These mechanisms can reduce interruption without eliminating the underlying movement of state. Our starting point is a deployment in which the main application objects can remain accessible across the move. This raises a different question: which state must be reconstructed when the objects themselves need not move?

\subsection{Memory placement as a migration decision}
CXL pooling and tiering systems separate memory placement from the compute node that uses it~\cite{pond,tpp}. For migration, the useful property is that both endpoints can access the same retained contents. This property is stronger than placing checkpoint files on shared storage: the running service directly reads and writes the objects that the destination will later use. It also differs from merely adding memory capacity to one node. The destination must reach the same backing region and observe the writes that precede takeover.

Object placement must precede migration to remove copying from the interruption path. Moving an ordinary heap into shared memory only after the source stops would still require copying that heap before execution could continue. \system{} instead makes shared placement part of allocation during normal execution. Ordinary memory remains available for stacks, runtime structures, and allocations outside the shared arena. Migration can then treat retained contents and ordinary process state differently without requiring every allocation to follow one path.

This choice makes shared-region availability a prerequisite for continuation. CXL access costs and failure behavior remain separate concerns~\cite{genuine,aifm,failure}. The prototype evaluates the software mechanism with emulated CXL memory; shared accessibility does not by itself provide power-loss persistence or a second copy of the service's state.

\subsection{Why shared bytes do not resume execution}
Native references connect objects across memory regions. Consider a stack variable that points to a shared hash entry, which in turn points to an ordinary-memory buffer. Restoring the shared region at another virtual address breaks the stack reference. Restoring it at the original address over another backing region preserves the pointer bits but changes the referenced object. Restoring only shared memory leaves the entry's reference to the ordinary buffer unresolved. Continuation therefore requires both memory domains and their original relationship.

The relationship is temporal as well as spatial. A captured stack cannot safely continue over a shared object that another writer has changed since the checkpoint. Freezing the source process tree establishes a common execution boundary for retained objects and captured process state. Keeping application writers inactive preserves that correspondence until the destination resumes. This is a process checkpoint boundary; it does not introduce an application transaction boundary or permit concurrent writers to proceed independently.

These dependencies motivate address preservation and coordinated handoff. A destination must reconstruct the captured ordinary state, reconnect the retained region using its saved mapping information, and restore supported resources before threads resume. Exposing the shared bytes is one part of this operation.

\subsection{Relationship to replication}
Redis replication maintains a separate logical dataset through synchronization and an update stream, then promotes a replica~\cite{redis-replication}. \system{} transfers execution over objects already in shared memory, including the captured allocator and process state needed to use them. It avoids maintaining another application-level dataset for this migration path and uses CRIU's resource reconstruction across the evaluated services. The trade-off is a pause for capture and takeover, together with continued dependence on the shared region. Replication preserves another copy and has its own synchronization and promotion protocol. The two approaches serve different preparation and failure assumptions; retained-object migration does not replace the redundancy needed to survive loss of shared state.

\section{Design}\label{sec:design}\label{sec:contract}
\subsection{Separate retained objects from captured execution}
\begin{figure*}[tbp]
\centering\includegraphics[width=\linewidth]{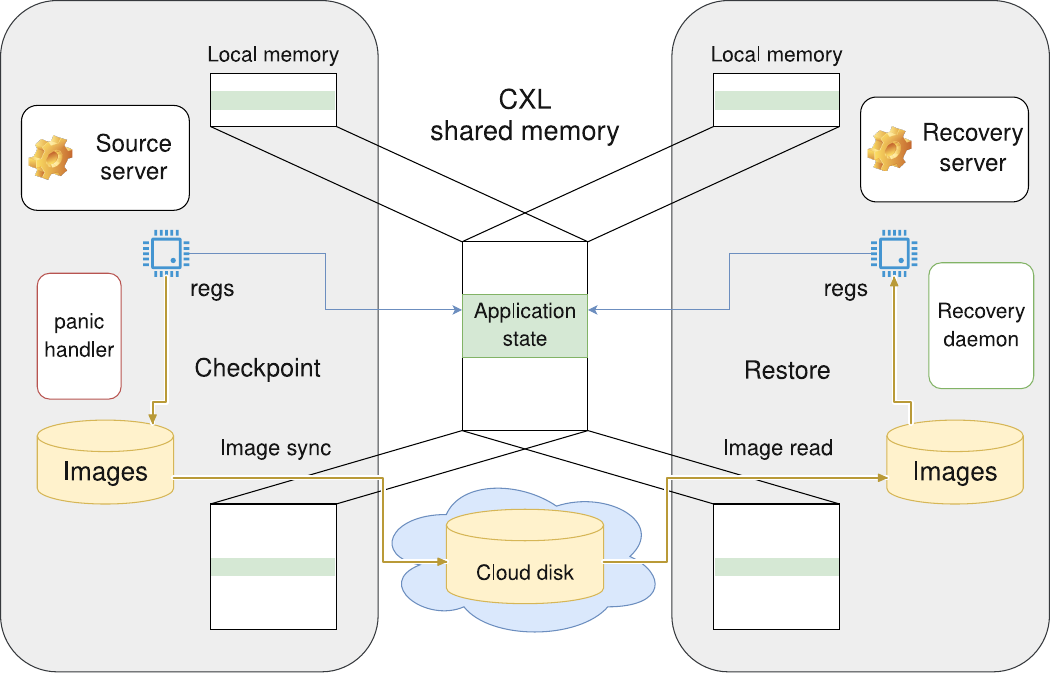}
\caption{\system{} architecture. Shared application objects remain in CXL memory; other process state and mapping records follow the checkpoint-storage path. The panic handler and recovery daemon depict a recovery extension. The evaluation studies controlled migration with an emulated CXL device.}
\label{fig:state}
\Description{Shared application objects connect the source and destination while other process state follows checkpoint storage.}
\end{figure*}

\system{} combines an allocation layer with a CRIU mapping extension (Figure~\ref{fig:state}). The allocation layer places eligible application objects in CXL memory, and CRIU captures the process state needed to continue using those objects. A panic handler and a destination recovery daemon connect checkpoint and restore to the kernel's panic path. These mechanisms allow the location of execution to change while the principal object contents remain in place.

The state partition follows actual memory placement. Contents of supported shared mappings remain in the backing region, while registers, stacks, ordinary pages, and supported resources enter checkpoint images. Mapping records associate the retained contents with their original process addresses. Allocator metadata belongs to the retained or captured part according to where it resides. This partition avoids assuming that all application state is shared or that all allocator state is ordinary memory.

The mapping is the classification unit, rather than an application object type. CRIU can distinguish the retained region without walking Redis or Memcached data structures or interpreting individual pointer fields. Objects outside that region continue through the ordinary checkpoint path. This permits a mixed address space while retaining CRIU's responsibility for the remaining process state.

\subsection{Establish residency before checkpointing}\label{sec:residency}
The allocation layer preserves the application's allocation interfaces and object representation. It acquires a shared region during execution and directs eligible allocations into it. The service consequently builds its data structures over addresses that can be recreated at the destination. Choosing this placement before migration removes a conversion of the live object graph from the handoff procedure.

An alternative is to copy or relocate objects into shared memory when a checkpoint is requested. That would retain the original allocation policy until migration, but the source would still need to traverse or copy its state on the migration path. Relocation would also have to preserve references held outside the moved region. Preexisting shared placement avoids this work, at the cost of requiring applications to use an allocation path that reaches the shared arena.

The design permits ordinary-memory fallback rather than treating partial shared residency as a different checkpoint format. If an allocation resides outside the arena, CRIU captures the necessary pages containing it. The benefit therefore depends on how much state remains in shared memory and on the ordinary memory and resources still requiring reconstruction. It does not follow from the size of a reserved shared region alone.

Allocator continuity is part of object continuity. A new allocator initialized over a populated region would not automatically know which blocks contain live objects. \system{} restores the existing allocator's control state and references along with application execution. The destination resumes management of the retained allocations instead of rebuilding an empty heap over them.

\subsection{Capture and publish one consistent checkpoint}\label{sec:publication}
Checkpointing first freezes the source process tree. With no other application writer modifying the shared region, the retained objects and serialized execution state correspond to the same instant. CRIU records the ordinary pages, mapping information, and supported resources while the retained objects remain accessible in their existing backing memory. The source stays inactive throughout destination reconstruction.

This ordering matters for references in both directions. Capturing a stack before an object mutation, but retaining the object's later contents, could resume execution against a state that the source never observed at that point. Freezing the whole process tree also captures the allocator and application threads together. The design assumes that writers outside the captured tree do not independently modify its retained objects during this interval.

Checkpoint publication coordinates the image path with restore. In the shared-storage deployment, a CRIU post-checkpoint callback synchronizes the generated checkpoint files and publishes a completion marker. The destination waits for that marker and for the required container metadata before starting restore. The marker establishes readiness of that checkpoint's files; visibility of the retained memory is a separate prerequisite. Synchronizing files does not create a persistent snapshot of the shared heap.

Local-image migration has an additional availability step: the checkpoint files must reach the destination before restore. Shared checkpoint storage removes that explicit transfer, while shared application memory removes the retained contents from page serialization and loading. These changes affect different parts of migration. The evaluation compares the resulting deployment configurations and distinguishes core checkpoint processing from client-visible interruption.

\subsection{Reconstruct mappings before resuming threads}\label{sec:mapping}
The destination reconnects each retained region using its captured virtual-address range, backing offset, protections, and shared attributes. It must attach the same backing region and observe the source's completed writes. The prototype associates device paths with regions through deployment configuration. Equal path strings on two endpoints are insufficient unless that configuration makes them refer to the same contents.

The initial allocation mapping may use an address chosen by the operating system. Restore reproduces the address that was captured, rather than requiring every service to start at a globally predetermined address. This preserves native pointers within shared memory and from ordinary memory into shared objects. References in the opposite direction additionally require CRIU to restore the ordinary pages and their virtual layout.

Returning to the stack, shared-entry, and ordinary-buffer example, the destination restores the stack and buffer from the checkpoint and maps the entry's retained contents at the saved address. Each pointer then denotes the same object as at capture. No application-specific pointer translation is needed. This argument depends on the target accepting the recorded mappings; a different address cannot be substituted transparently when a required mapping fails.

Application threads resume after mappings and supported resources have been reconstructed. Retained objects are addressable without a separate bulk-fetch phase for their contents. Ordinary page restoration, address-space construction, and resource handling still require work. This is why the design reduces state copying without equating remapping with an instantaneous service restart.

\subsection{Takeover and recovery scope}
The source container stops before the destination resumes from its checkpoint, leaving only the destination to modify the retained objects. Reconstructing the process and taking over its network identity then make the service reachable. This ordering assumes one designated destination. A checkpoint completion marker does not select among competing destinations or prevent another node from writing to the shared region; deployments with independent recovery decisions require an additional mechanism to exclude concurrent writers.

The panic handler provides a way to initiate this process when a source enters \code{panic()}. It attempts to create a checkpoint before the kernel stops the other CPUs. A recovery daemon at the destination waits for checkpoint completion and restores the container using the retained objects. This path requires scheduling, the container runtime, checkpoint storage, and the shared region to remain usable despite the triggering fault. It therefore targets faults that leave these facilities functional, rather than failures that immediately prevent further execution. The evaluation below uses planned migration and does not measure this panic-triggered path.

Pre-copy and post-copy optimize transfer while retaining some image-backed state movement~\cite{clark,m3u}. \system{} relies on prior placement to keep its shared objects accessible through the move. The destination does not defer retrieval of those objects until subsequent requests, but it remains dependent on the shared backend. This makes the design appropriate to migration within a shared-memory deployment rather than to moving a service to an arbitrary host without access to its retained state.

The retained heap is mutable, so its checkpoint association lasts only while the captured contents remain unchanged. Once the destination modifies an object, an older execution image no longer necessarily describes the corresponding heap. Another migration establishes a new freeze point. Historical rollback, competing-writer recovery, and survival of shared-region loss require state preservation or coordination beyond this mechanism.

\section{Implementation}\label{sec:implementation}
\subsection{Shared allocation and allocator state}
\begin{figure*}[tbp]
\centering\includegraphics[width=.78\linewidth]{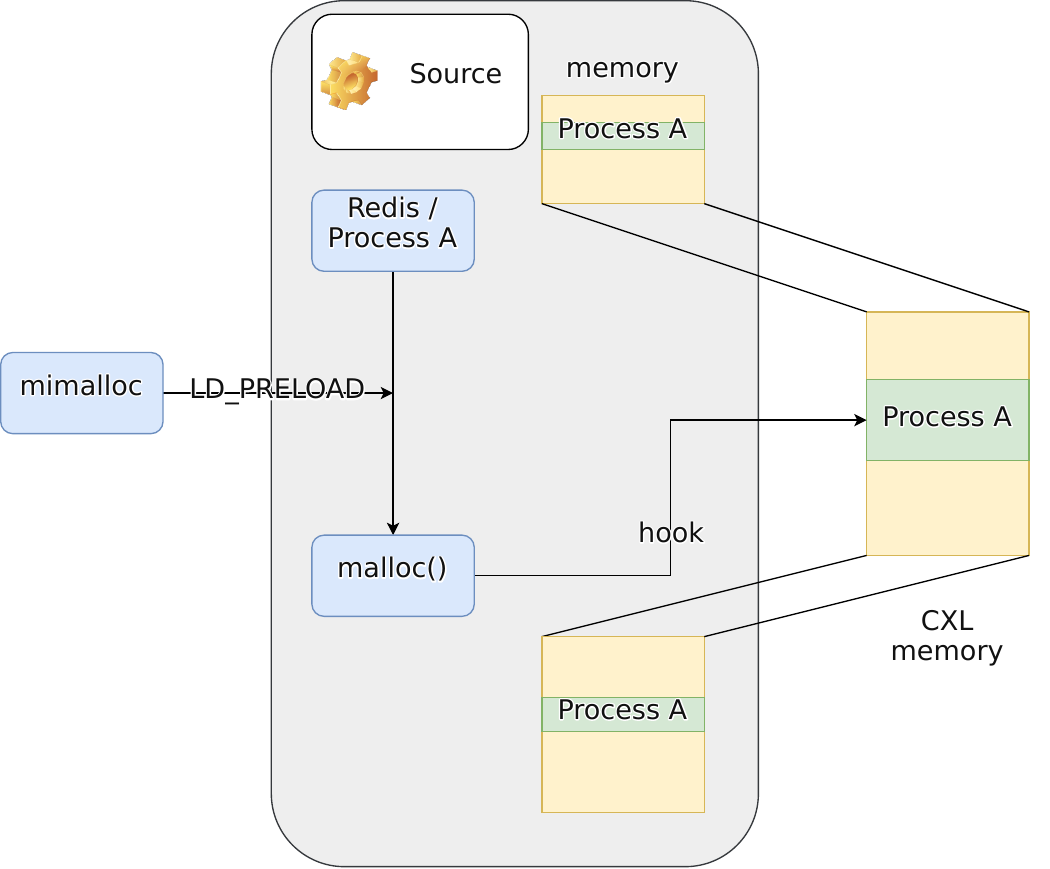}
\caption{Shared allocation in \system{}. Allocation interception places eligible objects in CXL memory; checkpoint and restore preserve their mappings and allocator state.}
\label{fig:allocation}
\Description{An allocation interceptor routes eligible allocations into shared memory and preserves their mappings across migration.}
\end{figure*}

The allocation component is a C library loaded through \code{LD_PRELOAD}. Redis uses a libc-allocation build so that its requests reach the interposed interfaces. The component opens the CXL memory device, creates a readable/writable \code{MAP_SHARED} mapping, and registers an aligned part of it as a managed mimalloc 2.1.7 arena~\cite{mimalloc}. Initialization runs once per process, and per-thread heaps allocate from that arena (Figure~\ref{fig:allocation}).

The mapped address range determines ownership. Allocation requests use the shared arena, while deallocation and reallocation first distinguish shared pointers from libc pointers. A shared pointer returns to mimalloc; an ordinary pointer returns to libc. This routing matters because the process can contain both kinds of allocation, including objects created outside the intercepted path. Passing a shared pointer to the ordinary allocator would not preserve its allocation metadata.

Initial arena setup and later allocation failure follow different paths. Failure to open, map, or register the region prevents shared initialization. After initialization, unsuccessful shared \code{malloc} and \code{calloc} requests can fall back to libc. Growing a shared allocation uses mimalloc and reports failure if it cannot satisfy the request, leaving the original allocation available. These cases explain why the checkpoint follows actual mappings instead of inferring shared residency from the allocation interface alone.

Arena bounds, initialization state, arena identifiers, and thread-local heap references are part of the captured process. Metadata located in ordinary memory follows CRIU's page path; metadata located in the arena remains with the shared objects. Restore continues these structures at their original addresses. It does not rerun arena initialization over live allocations. This preserves the relationship needed by subsequent allocation, reallocation, and deallocation.

\subsection{CRIU device admission and mapping reconstruction}
The plugin extends CRIU 3.19's device-mapping interface. During memory-map inspection, it admits character-device mappings whose device identity matches an exposed CXL memory device. Other device mappings remain subject to CRIU's existing support checks. Admission lets the supported region use the file-backed shared-mapping path; it does not make arbitrary device state migratable.

CRIU's native mapping records retain the virtual range, backing offset, permissions, and sharing attributes. The retained contents stay outside ordinary page-image processing, while required anonymous and other ordinary pages continue through existing CRIU mechanisms. During restore, the file-handling path reopens the backing device and the restorer recreates the saved mapping with \code{MAP_FIXED}. The plugin thus supplies device recognition while CRIU preserves the address-space and resource relationships around that device.

An open device descriptor and an existing memory mapping have different lifetimes. The allocation library closes its initialization descriptor after creating the mapping, which remains valid. Native mapping records support reconstruction of that region. If a process retains an open supported device descriptor, additional plugin callbacks save and restore its device-path information in a versioned descriptor record. That record identifies how to reopen the descriptor; it does not contain the shared heap or replace the mapping metadata.

This division keeps the extension focused on the retained-memory path. Process-tree reconstruction, thread contexts, ordinary memory, namespaces, and supported descriptors and sockets remain CRIU responsibilities. The destination still needs a compatible execution environment and access to referenced files. Sharing the application heap does not remove those dependencies.

\subsection{Checkpoint triggering and service recovery}
\system{} uses Docker's checkpoint interface with CRIU 3.19. A compatible destination container is prepared in a stopped state, with the application image and shared-memory device available. The source checkpoint captures the running container's execution rather than starting a fresh application and importing its logical dataset. This permits the destination to resume the recorded process and allocator state together.

The prototype invokes the panic handler near the start of \code{panic()}, before the kernel's CPU-stop sequence. Through synchronous user-mode helpers, the handler requests a Docker checkpoint, copies the resulting images to shared storage, synchronizes them, and publishes a completion marker. The destination recovery daemon waits for that marker, checks the checkpoint images and required container metadata, and starts the stopped container from the checkpoint. This implementation depends on the kernel services identified in the recovery scope above.

The planned migrations evaluated below initiate checkpointing through migration scripts instead of the panic handler. With shared checkpoint storage, the destination independently watches the checkpoint directory. The post-checkpoint callback synchronizes generated files and publishes its completion marker only after that synchronization. Container-runtime metadata may become available separately, so the destination also waits for the required configuration and descriptor records. Restore begins when both the checkpoint and the container metadata are available; it does not depend on a separate source-issued restore request. Local-image migration transfers its files before the corresponding restore step.

In the continuous-request experiments, network activation follows process restoration. The destination initially avoids claiming the service's original link-layer identity. After the process and its supported sockets have been restored, the migration scripts activate the original network identity and announce it to the local network. This sequencing keeps traffic directed away from an unprepared destination and accounts for interruption beyond CRIU's core phases.

The allocation integration targets applications whose requests reach the interposed interfaces. Statically linked programs and runtimes with independent allocators would require a separate integration path. The implementation also relies on configured access to one shared region and a single destination writer. The QEMU setup supplies those memory-access conditions for functional evaluation. The prototype does not implement distributed failure detection or a fencing protocol.

\section{Evaluation}\label{sec:method}\label{sec:results}
\subsection{Experimental setup}
We evaluate two deployment configurations in virtual machines on one host. QEMU exposes the same host-DRAM-backed region as an 8~GiB CXL Type-3 device to both endpoints~\cite{qemu}. The virtual machines run openEuler 24.03 LTS-SP1 with Linux 6.6 and Docker 17.06.2-ce. Both configurations use CRIU 3.19, Redis 5.0.14, and Memcached 1.6.31. CRIU uses ordinary memory, libc allocation, and local ext4 checkpoint storage; \system{} uses shared CXL memory, mimalloc, and shared NFS~4.2 storage.

Dataset size denotes the total bytes in values or structure members, excluding keys and metadata. We vary datasets from 1--6~GiB and use a separate 3~GiB dataset for continuous GET, SET, and equal read/write workloads. Constructed data structures and states derived from Twitter and Meta cache workloads broaden allocation coverage~\cite{twitter,twitter-trace,meta-trace}. Trace replay preserves operation order, value lengths, and logical expiration times using synthetic values and anonymized keys. Replay prepares the state before migration; Meta lease operations map to GET and SET without implementing the lease protocol.

We measure elapsed CRIU checkpoint and restore time, including checkpoint synchronization. Their sum excludes migration steps outside CRIU, such as image transfer and network takeover. Client-visible interruption includes these additional stages. The comparison evaluates the complete deployment: allocator, checkpoint storage, and available memory capacity differ together. The emulated device supports functional evaluation rather than physical CXL performance measurement.

\begin{figure*}[tbp]
\centering\includegraphics[width=\linewidth]{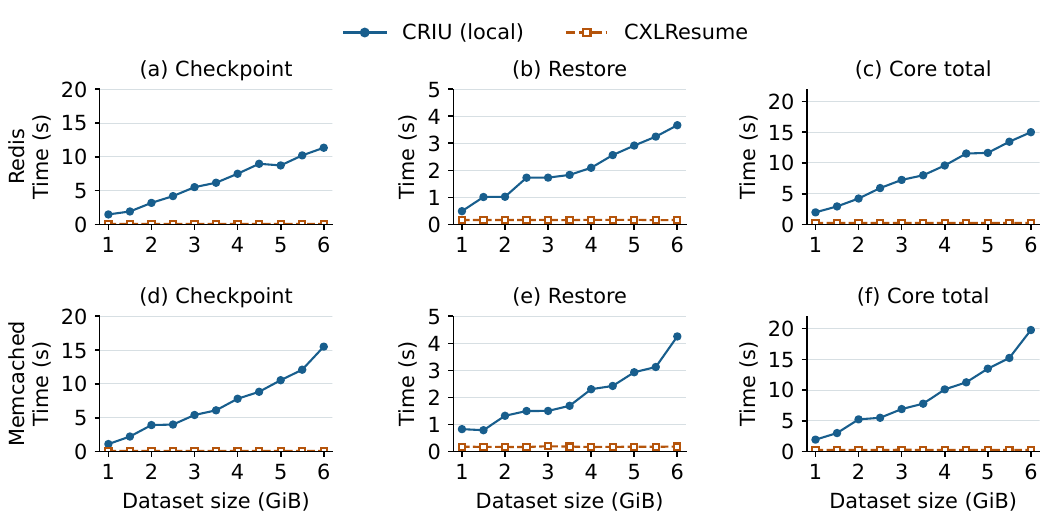}
\caption{Checkpoint and restore time versus dataset size. Time axes are linear and start at zero, with the same scale for each metric across services. CRIU uses local images; CXLResume uses shared images.}
\label{fig:capacity}
\Description{Line charts compare checkpoint, restore, and total core time across dataset sizes for both services and deployment configurations.}
\end{figure*}

\begin{figure*}[tbp]
\centering\includegraphics[width=\linewidth]{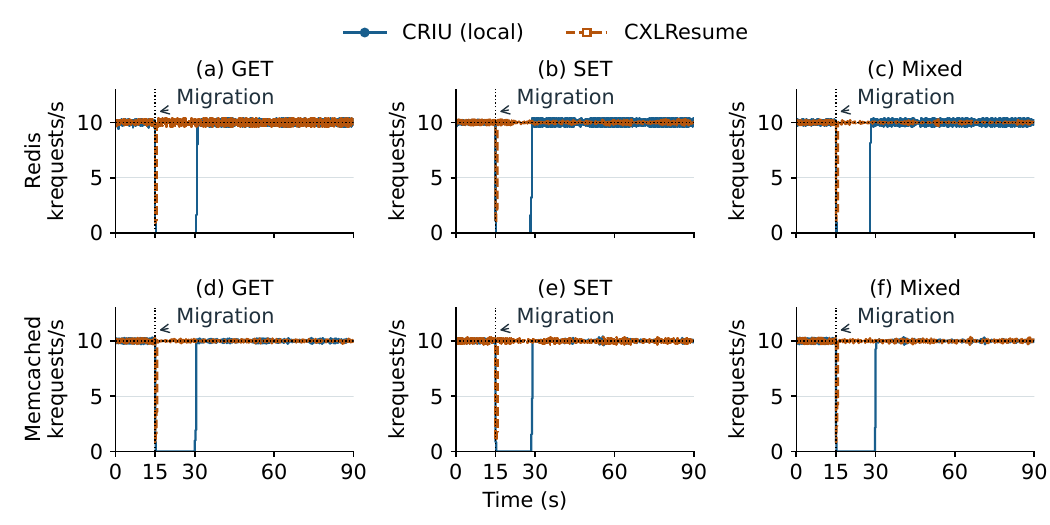}
\caption{Successful request completions during the 90~s workload, in 0.25~s bins. Time zero is workload start; migration begins near 15~s. Each curve is a selected migration. Dotted horizontal lines mark the 10,000-request/s cap.}
\label{fig:timeline}
\Description{Request-completion traces show steady service, migration interruption, and resumed service under read, write, and mixed workloads.}
\end{figure*}

\begin{figure*}[tbp]
\centering\includegraphics[width=\linewidth]{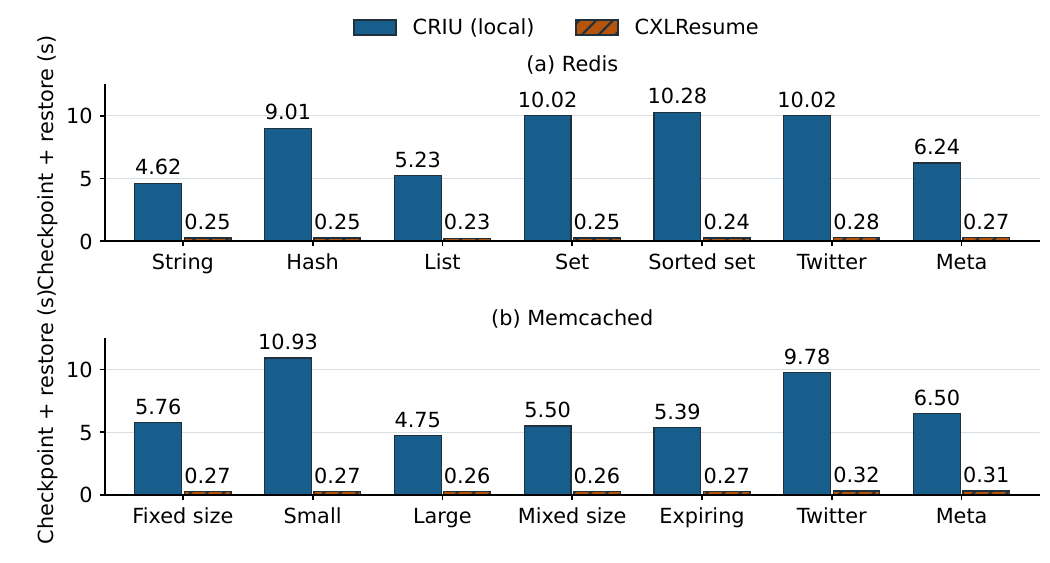}
\caption{Checkpoint and restore across application states, on a common zero-origin linear scale. Most states contain 2.5~GiB of value or member data; Meta-derived states target 2~GiB.}
\label{fig:shape}
\Description{Grouped bars compare checkpoint and restore time across structured and trace-derived application states.}
\end{figure*}

\subsection{Checkpoint cost as the dataset grows}\label{sec:capacity}
Figure~\ref{fig:capacity} compares checkpoint and restore as the dataset grows. Retaining application objects limits the amount of state that must pass through CRIU's page-image path.

At 6~GiB, CRIU's combined checkpoint and restore time is 15.015~s for Redis and 19.770~s for Memcached, compared with 0.242~s and 0.276~s for \system{}. CRIU's core time is therefore 61.98 and 71.56 times that of \system{}, respectively. The retained-object path reduces the growth in checkpoint processing while preserving the ordinary-state work required to reconstruct the process.

\subsection{Service interruption under continuous requests}\label{sec:gaps}
To relate checkpoint processing to service continuity, we drive a continuous workload with memtier 2.2.0~\cite{memtier}. A single TCP connection with one outstanding request is capped at 10,000 requests/s. The workload lasts 90~s, with migration initiated near 15~s. The client remains at the destination, accessing the remote source before migration and the local service afterward.

We define the migration interval from the trigger to observed restore completion. The response gap is the longest interval between consecutive successful responses that overlaps this period, measured using the client's monotonic clock. Figure~\ref{fig:timeline} uses one trace per configuration, selected for the closest match between this response gap and the checkpoint, image-transfer, and restore total. These traces illustrate the recovery sequence rather than migration variability.

The traces show steady request processing before migration, an interruption during takeover, and resumed processing at the destination. In the selected GET traces, \system{} has response gaps of 0.396~s for Redis and 0.410~s for Memcached; CRIU's corresponding gaps are 15.540~s and 15.214~s. SET and mixed workloads show the same ordering. Both configurations return close to the configured request rate, so the observed benefit is earlier resumption.

Response gaps exceed core checkpoint and restore time because they also include work outside CRIU, such as image transfer and network takeover. Retaining objects shortens the state-processing stages, while the client measurement captures when that reduction becomes visible to the service.

\subsection{Continuation across application states}\label{sec:shape}
We evaluate Redis String, Hash, List, Set, and Sorted Set states, and Memcached states with fixed-size, small, large, mixed-size, and expiring objects. Each service also uses Twitter- and Meta-derived states. Figure~\ref{fig:shape} compares checkpoint and restore across these layouts.

The two configurations use the same logical dataset within each workload. Across workloads, keys, indexes, and allocation metadata change the process state associated with a given value-data size. Meta-derived states also use a smaller dataset. Checkpoint cost therefore depends on state layout as well as value-data size.

Across the evaluated states, CRIU requires seconds for checkpoint and restore, while \system{} completes the core phases in less than one second. Sampled values and data-structure fingerprints remain consistent after migration, accounting for expired entries. Together, these results support service continuation across the tested object layouts.

\FloatBarrier
\section{Related work}\label{sec:related}
\paragraph{Checkpoint placement and migration.}
Live migration distinguishes service downtime from total migration time~\cite{clark}, and post-copy systems such as M3U optimize memory transfer~\cite{m3u}. VAS-CRIU uses copy-on-write snapshots with CRIU resource capture~\cite{vas}; fabric-attached post-copy uses shared memory for checkpoint and page-transfer state~\cite{fam}. \system{} instead places writable application objects in shared memory before migration and coordinates their continued use with captured execution.

\paragraph{Execution reuse and remote fork.}
Snapshot-loading systems reduce startup work through memory and operating-system reuse~\cite{reap,faasnap,spice}. TrEnv shares execution environments across functions and nodes~\cite{trenv}; Mitosis and CXLfork support remote process cloning using RDMA and CXL, respectively~\cite{mitosis,cxlfork}. Their reuse mechanisms create instances. \system{} transfers a service's writable use of existing objects while preserving references across shared and ordinary memory.

\paragraph{Shared-memory recovery.}
Shared allocation and recovery systems address reference management and partial failures~\cite{cxlshm,cxlalloc}; persistent-memory systems address relocation and coordinated checkpoints~\cite{puddles,treesls}. CXL failure models and checking tools make ordering and failure assumptions explicit~\cite{failure,cxlmc}, while Tigon studies CXL shared memory for databases~\cite{tigon}. These mechanisms complement process continuation by addressing allocation, persistence, and coordination.

Other systems recover through redundancy at the shared-memory, object, remote-memory, or storage layer~\cite{oak,apta,hydra,carbink,ramcloud}. Remus coordinates virtual-machine replication with external output~\cite{remus}. \system{} assumes continued access to retained objects; it does not supply the redundancy needed to survive loss of that state.

\paragraph{Live update.}
Ksplice and Kitsune preserve service while changing kernel or application code~\cite{ksplice,kitsune}. \system{} addresses a change in placement while preserving the executable and captured state. Combining migration with code or schema evolution remains outside the evaluated mechanism.

\section{Discussion}\label{sec:limits}
The measurements compare complete deployments whose allocators, checkpoint storage, and available memory capacity differ. They do not isolate retained-object checkpointing from storage I/O or memory-pressure effects. Isolating these factors would require matched allocators, checkpoint storage, and memory budgets.

The service experiments use a paced closed-loop connection, and the selected traces illustrate recovery rather than its variability. High-concurrency queuing and tail latency remain unmeasured. Content validation is limited to sampled values and structure fingerprints, and trace-derived states broaden allocation coverage without reproducing production workloads.

The prototype also assumes that both endpoints can access the same shared region and that only one writer resumes. A single-host, DRAM-backed emulation does not establish physical CXL performance or survival of power loss, backend loss, arbitrary source crashes, or partitions with competing writers. Extending the mechanism to those failures requires shared-region visibility and exclusive takeover guarantees~\cite{failure,cxlmc}.

\section{Conclusion}\label{sec:conclusion}
\system{} combines shared allocation with address-preserving checkpoint and restore so that execution can move while application objects remain in place. Redis and Memcached experiments on DRAM-backed CXL emulation show shorter core phases and subsecond response gaps in the evaluated service traces. Retained-object migration reduces the state copied during takeover while relying on shared-memory availability and a single destination writer.

\FloatBarrier
\clearpage
\bibliographystyle{ACM-Reference-Format}
\bibliography{references}

\FloatBarrier
\end{document}